# Artificially engineered superlattices of pnictide superconductor

S. Lee[1], C. Tarantini[2], P. Gao[3], J. Jiang[2], J. D. Weiss[2], F. Kametani[2], C. M. Folkman[1], Y. Zhang[3], X. Q. Pan[3], E. E. Hellstrom[2], D. C. Larbalestier[2], and C. B. Eom[1*]

[1]Department of Materials Science and Engineering, University of Wisconsin-Madison, Madison, WI 53706, USA

[2]Applied Superconductivity Center, National High Magnetic Field Laboratory, Florida State University, 2031 East Paul Dirac Drive, Tallahassee, FL 32310, USA

[3]Department of Materials Science and Engineering, The University of Michigan, Ann Arbor, Michigan 48109, USA

**Significant progress has been achieved in fabricating high quality bulk and thin-film iron-based superconductors. In particular, artificial layered pnictide superlattices[1,2] offer the possibility of tailoring the superconducting properties and understanding the mechanism of the superconductivity itself. For high field applications, large critical current densities ($J_c$) and irreversibility fields ($H_{irr}$) are indispensable along all crystal directions. On the other hand, the development of superconducting devices such as tunnel junctions requires multilayered heterostructures. Here we show that artificially engineered undoped Ba-122 / Co-doped Ba-122 compositionally modulated superlattices produce *ab*-aligned nanoparticle arrays. These layer and self-assemble along c-axis aligned defects[3–5], and combine to produce very large $J_c$ and $H_{irr}$ enhancements over a wide angular range. We also demonstrate a structurally modulated SrTiO$_3$ (STO) / Co-**

**doped Ba-122 superlattice with sharp interfaces. Success in superlattice fabrication involving pnictides will aid the progress of heterostructured systems exhibiting novel interfacial phenomena and device applications.**

Since the discovery of iron-based superconductors[6–11], epitaxial thin films have been successfully grown by many groups[3,12–18] and they have significantly advanced potential device applications and the understanding of the fundamental physical properties of these new superconductors.[12,19–25] In particular, we reported high quality Co-doped $BaFe_2As_2$ (Ba-122) single crystal thin films using template engineering which generated $c$-axis aligned, self-assembled, second phase nanorods[3–5] as schematized in Fig. 1a. These nanorods yielded strong $c$-axis pinning centres that enhance the in-field critical current density, $J_c$. The $c$-axis pinning was so strong that actually it was able to invert the usual irreversibility field anisotropy in which $H_{irr}$ parallel to the $ab$-plane is 1.5-2 times that parallel to the $c$-axis. This motivated us to consider using multilayer growth methods to enhance the $ab$-plane properties since high field applications prefer to enhance $J_c$ in all field configurations, as sketched in Fig. 1c. We note that no study of artificially controlled flux pinning for pnictide superconductor has yet been reported, while many investigations of $J_c$ and irreversibility field ($H_{irr}$) enhancement by nanoparticle incorporation have been performed in the high $T_c$ cuprate superconductor $YBa_2Cu_3O_{7-x}$ (YBCO) like that using $YBCO/Y_2BaCuO_5$ and $YBCO/BaZrO_3$ multilayers[26,27]. But every material system is different and it is striking that pinning engineering has not yet been demonstrated in either the Bi-2212 or Bi-2223 systems or even the most classical high field superconductor, $Nb_3Sn$, even though it is very possible and valuable in YBCO. This motivation underpinned our studies of this Fe-based superconductor system. A major difference between YBCO and Fe-based superconductor systems is that Fe-based superconductors are metallic and it thus seemed questionable whether insulating epitaxial oxide pinning centers grown parallel to the *ab*-planes could be incorporated without suppressing the superconductivity of the Ba-122 matrix. We were at first doubtful that multiple epitaxial layers of a Ba-122 metal phase and an oxide phase could be grown with high crystalline perfection since they might have been structurally and chemically incompatible but in fact we here demonstrate that structurally and compositionally modulated superlattices of oxide can be grown in the matrix of a Ba-122 metallic system, a result which is significantly different from the YBCO oxide superconductor system, where both the oxide matrix and the pinning phase are structurally and chemically compatible. We see that this enables strong vortex pinning and opens up the possibility of growing superconducting superlattice structures (schematized in Fig. 1b) that can be used as a means for understanding fundamental mechanisms of superconductivity and development of superconducting devices such as pnictide tunnel junctions.

Taking account of the metallic properties of Ba-122 which does not contain oxygen and is easily compromised by impurities, we narrowed down the possible interlayer to two different types of model superlattice structure. Since in our previous work[3], we found that a divalent, alkaline earth element-containing oxide like $SrTiO_3$ (STO) is an excellent platform on which to grow Co-doped Ba-122 thin films due to their common features, STO was our first choice. This led to a structurally modulated superlattice with sharp interfaces, devoid of *c*-axis defects. The second choice of interlayer was undoped Ba-122 made from an oxygen-rich, Co-free 122 target (O-Ba-122) where the interlayer has a similar structure but slightly different lattice parameter. This led to compositionally modulated O-Ba-122 / Co-doped Ba-122 superlattice. Our hypothesis was that oxygen-rich undoped Ba-122 would facilitate the formation of strong pinning, oxygen-rich precipitates along the *ab*-planes, just as it had for the *c*-axis pins in single layer films of Co-doped Ba-122.

With this approach, we have successfully grown epitaxial superlattices of $STO_{1.2\ nm}$ / 8 % Co-doped Ba-122$_{13\ nm}$ and O-Ba-122$_{3\ nm}$ / 8 % Co-doped Ba-122$_{13\ and\ 20\ nm}$ on 40 nm STO templates deposited on (001) (La,Sr)(Al,Ta)O$_3$ (LSAT) single-crystal substrates by using pulsed laser deposition. We varied the thickness of the bilayer (modulation wavelength = $\Lambda$) and the total number of bilayers (n). Their structural and superconducting properties are listed in Table SI in supplementary information. To focus on the key points, we discuss here only ($STO_{1.2\ nm}$ / Co-doped Ba122$_{13\ nm}$) × (n=24) superlattice (hereafter referred to as STO SL) and (O-Ba-122$_{3\ nm}$ / Co-doped Ba122$_{13\ nm}$) × (n=24) superlattice (hereafter referred to as O-Ba-122 SL).

The epitaxial crystalline quality and modulation wavelength ($\Lambda$) of the superlattices were determined by four-circle x-ray diffraction (XRD) with a Cu K$_\alpha$ source ($\lambda$ = 1.5405 Å). Figure 2a shows the $\theta$-$2\theta$ scan of the STO SL. The XRD pattern shows that Co-doped Ba-122 00*l* reflections dominate, which indicates *c*-axis epitaxial growth normal to the STO template and LSAT substrate. Figure 2c is a magnification of Fig. 2a close to the 002 reflection of the STO SL which clearly shows satellite peaks with calculated modulation length $\Lambda$ = 14 nm ± 2 nm the same as the nominal $\Lambda$. Rocking curves for the 004 reflection were measured to determine the out-of-plane mosaic spread and crystalline quality. As shown in Fig. 2d, the full width at half maximum (FWHM) of the 004 reflection rocking curve of STO SL is 0.97°. Furthermore, the STO SL exhibits strong and sharp peaks ($\Delta\phi$ = 1.1°) only every 90° in the azimuthal phi scan of the off-axis 112 reflection of Co-doped Ba-122 (Fig. S1a) revealing a good in-plane epitaxy. These results confirm the excellent epitaxial arrangement, even though as many as 24 STO / Co-doped Ba-122 bi-layers were grown.

Figure 2b is the $\theta$-$2\theta$ scan of the O-Ba-122 SL and it shows excellent $c$-axis epitaxial growth. Fig. 2d shows the FWHM of the 004 reflection rocking curve of O-Ba-122 SL to be as narrow as 0.26°, which is far superior to the STO SL and even better than other reports about single layer Ba-122 thin films[13,16]. Furthermore, as shown in Fig. S1, the azimuthal phi scan of O-Ba-122 SL shows much sharper peaks ($\Delta\phi = 0.69°$) than the STO SL. The reason why O-Ba-122 SL has superior crystalline quality is that O-Ba-122 and Co-doped Ba-122 are structurally identical, differing only by the 8% cobalt addition. Since the atomic scattering factors of Co and Fe are very close it would be exceedingly difficult to see any satellite peaks by x-ray diffraction.

To investigate the microstructure of STO and O-Ba-122 SLs, transmission electron microscopy (TEM) was used. Figure 3a, b show cross sectional low and high magnification high-angle annular dark field (HAADF) images of STO SL. In Fig. 3a, bright and dark layers correspond to 13 nm Co-doped Ba-122 layer and 1.2 nm STO layer, respectively. We can clearly see that there are 24 STO / Co-doped Ba-122 bi-layers and $\Lambda$ is 14 nm in accordance with our design and the modulation wavelength determined by x-ray diffraction. Furthermore, 3 unit-cell-thick STO interlayers have been uniformly grown on the Co-doped Ba-122 layer as shown in Fig. 3b. This indicates that we can control the thickness of the STO interlayer with single unit-cell precision and maintain a sharp interface between STO and Co-doped Ba-122. A schematic of the <100> projection of Co-doped Ba-122 and STO at the interface is shown to the right of Fig 3b, where the bonding of the FeAs layer to the SrO layer in STO is quite evident, in excellent agreement with the hypothesis we previously proposed[3].

Figure 3c is a cross-sectional TEM image of the O-Ba-122 SL, which clearly shows 24 bilayers and the modulation wavelength $\Lambda=16$ nm. The planar view TEM images (Figure 3d) exhibits $c$-axis aligned defects as we observed in the single layer Co-doped Ba-122 in our previous reports[4,5]. The first interesting point is that the interface O-Ba-122 layers have grown as laterally aligned but discontinuous second phase nano particles of several nm size. Presumably, island growth of the oxygen-containing second phase is energetically favourable due to the small lattice mismatch with the Co-doped Ba-122 SL. Similar discontinuous growth was observed with $YBa_2Cu_3O_7$/ $YBa_2CuO_5$ multilayers[26]. The second interesting point is that the O-Ba-122 SL has *c*-axis defects extending across many *ab*-axis aligned nano particle arrays, while the STO SL does not have such *c*-axis aligned defects. We infer that oxygen which is needed to produce the *c*-axis aligned defects is absorbed by the STO interlayers under our high-vacuum, high-temperature growth conditions. The structure of the O-Ba-122 SL is thus akin to Fig. 1c which is more desirable than the structure of the STO SL (Fig. 1b) for flux pinning, since the O-Ba-122 SL has structural defects in both *c*- and *ab*-axis directions. Strictly speaking, the actual morphology of O-Ba-122 SL is close to Fig. 1c with nanoparticles rather than a continuous bilayer interface.

To characterize the superconducting transition temperature $T_c$, resistivity was measured as a function of temperature ($\rho$-T) by the van der Pauw method. As shown in Fig. 4 the residual resistivity ratio (RRR) of the O-Ba-122 SL (1.7) is slightly lower than that of the STO SL (2.2). In the inset of Fig. 4, O-Ba-122 SL shows high $T_{c,\,\rho=0}$ = 22.9 K, and narrow $\Delta T_c$ of 1.1 K, which indicates the good quality of these layered films even in the presence of significant lateral second phase. By contrast, $T_{c,\,\rho=0}$ of the ultrathin layers of STO SL is suppressed to 17.0 K. Evidently, compatibility between the O-Ba-122 and Co-doped Ba-122 is better than that between STO and Co-doped Ba-122. Figure 4 also indicates a $T_{c,\,\rho=0}$ of 20.5 K for ~400 nm thick single layer Co-doped Ba-122 film[5] deposited on 40 nm STO/LSAT.

In order to understand the effect of this nanostructural engineering on $J_c$ and $H_{irr}$ we made extensive characterizations at various temperatures, fields and field orientations to the crystal axes. Figure 5a shows $J_c(H)$ for H//c at 4.2K far from $T_c$. It is immediately clear that the two samples with *c*-axis nanorod pinning defects have much higher $J_c(H)$ and $H_{irr}(T)$. The influence of the *ab*-plane defects is revealed in Fig. 5b where the $J_c$ anisotropy for the O-Ba-122 SL and Co-doped Ba-122 single layers at 16K for perpendicular (H//c) and parallel (H//ab) configurations is compared. Despite these data being affected by their $T_c$ differences (the O-Ba-122 SL has a higher $T_c$, consistent with its higher $J_c$), both samples have the same $H_{irr}$~11T for H parallel to the *c*-axis, a result consistent with the lower density of *c*-axis pinning centres in O-Ba-122 SL compared to the single layer Co-doped Ba-122. But the really striking result is that the inverted $H_{irr}$ anisotropy seen for the single layer film with only *c*-axis defects ($H_{irr}$ for H//ab is less than for H//c) is corrected when *ab*-plane pins are present in the O-Ba-122 SL film. It is clear that $H_{irr}$ for H//ab ($H_{irr,ab}$) of O-Ba-122 SL is approximately doubled from 9 to ~19 T, restoring the expected anisotropy of $H_{irr}$ and $J_c(H)$ without any degradation to the *c*-axis properties. This enhancement is due to the presence of the *ab*-plane aligned nanoparticles in the O-Ba-122 SL shown in Fig. 3.

The angular transport $J_c$ of the STO SL, O-Ba-122 SL, and Co-doped Ba-122 single layer[3–5] shown in Fig. 5c evaluated at a constant reduced temperature $T/T_c$ ~0.6 provides further insight into the pinning effects of the nanoparticles. The Co-doped Ba-122 single layer[3–5] shows only the strong *c*-axis pinning produced by the correlated, self-assembled nanopillars, while the $J_c$ of the STO SL shows only a sharp, few-degrees wide peak when the magnetic field is aligned with the *ab*-plane STO superlattice. The $J_c(\theta)$ of O-Ba-122 SL is higher than the other two samples and shows both strong *ab*-plane and *c*-axis peaks, which is quite consistent with the *ab*- plane aligned second-phase nanoparticles of the bilayer and its *c*-axis aligned defects, seen in Fig. 3c and d. Very strong pinning in fields up to 45 T has been reported elsewhere[28].

In summary, we have grown artificially layered superlattice structures in Co-doped Ba-122 thin films with controlled structural and compositional modulations. The insertion of O-Ba-122 layers allows nanoparticle formation that introduces strong vortex pinning along the *ab*-planes while still allowing the formation of vertically-aligned defects. The remarkable enhancement of the pinning properties over a wide angular range related to the *ab*-plane nanoparticles is highlighted by the significant increase of the irreversibility field and by much improved $J_c$. The engineered structures presented here are surely capable of further refinement by optimizing interlayer separation and the composition of the vortex pinning layers features which cannot be obtained in single layer films. The successful growth of such high quality artificially layered structure will have wide implications for achieving new interface-driven high $T_c$ superconductivity[29] and potential device applications involving SNS and SIS junctions[30]. Furthermore, artificially made multilayer structures can also be used as model systems to study many physical phenomena such as dimensionality, proximity effect and interface pinning.

**Methods**

(1.2 nm STO or 3 nm O-Ba-122 / 13 nm and 20nm Co-doped Ba-122) × n superlattice thin films were grown *in-situ* on STO - templated (001) oriented LSAT single-crystal substrates using pulsed laser deposition (PLD) with a KrF (248 nm) UV excimer laser in a vacuum of $3 \times 10^{-4}$ Pa at 730 ~750 °C. The base pressure before deposition was $3 \times 10^{-5}$ Pa, and the deposition took place at $3 \times 10^{-4}$ Pa due to degassing of the substrate heater. Laser triggering, ablation time, and target rotation were automatically controlled by computer program during the superlattice deposition. It took 5 seconds to change the target position and laser triggering between the Co-doped Ba-122 and O-Ba-122. The magnetization of films that were about 2 mm x 4 mm was measured in a 14 T Oxford vibrating sample magnetometer (VSM) at 4.2K with the applied field perpendicular to the film surface. For a thin film, $J_c = 15\Delta m/(V\,r)$, where $\Delta m$ is the width of the hysteresis loop in emu, V the film volume in cm$^3$, and r the radius corresponding to the total area of the sample size, and was calculated from $\pi r^2$ = a x b (a and b are the film width and length in cm, respectively).

**Acknowledgments**:

Work at the University of Wisconsin was supported by funding from the DOE Office of Basic Energy Sciences under award number DE-FG02-06ER46327. The works at the NHMFL was supported under NSF Cooperative Agreement DMR-0084173 and DMR-1006584, by the State of Florida and by AFOSR under grant FA9550-06-1-0474. TEM work was carried out at the University of Michigan and was supported by the Department of Energy under grant DE-FG02-07ER46416. We would like to thank Dillon Fong and Jenia Karapetrova and the APS for synchrotron experiments. S.L and C.B.E. would like to thank Satyabrata Patnaik for helpful discussions.

**Author Contributions**

S.L. fabricated Ba-122 superlattices, analysed epitaxial arrangement by x-ray diffraction and prepared the manuscript. C.T. carried out electromagnetic characterization and prepared the manuscript. P.G. and F.K and Y.Z. carried out TEM measurements. J.D.W. fabricated Ba-122 pulsed laser deposition targets for thin film deposition. J.J. carried out electromagnetic characterizations. C.B.E., D.C.L., E.E.H. and X.Q.P. supervised the experiments and contributed to manuscript preparation. C.B.E. conceived and directed the research. All authors discussed the results and implications and commented on the manuscript at all stages.

**Competing financial interests**

The authors declare no competing financial interests.

**Figure Captions**

**Figure 1.** **Schematics of various structures of superconductor epitaxial thin films. a**, superconductor thin film with defects along $c$-axis **b**, superlattice structure ($d_i$ = thickness of an insertion layer, $d_s$ = thickness of a superconductor layer, $\Lambda$ (modulation wavelength) = $d_i + d_s$, n = total number of bilayers) **c**, superlattice structure with defects along $c$-axis

**Figure 2.** **XRD patterns obtained on O-Ba-122 inserted and STO inserted Co-doped BaFe$_2$As$_2$ superlattices a**, Out-of-plane $\theta$-$2\theta$ XRD pattern of (STO$_{1.2\ nm}$/Co-doped Ba-122$_{13\ nm}$)×24. **b**, Out-of-plane $\theta$-$2\theta$ XRD pattern of (O-Ba-122$_{3\ nm}$/Co-doped Ba-122$_{13\ nm}$)×24. **c**, Magnified at near 002 reflection of (STO$_{1.2\ nm}$/Co-doped Ba-122$_{13\ nm}$)×24. **d**, Rocking curves and FWHM for (004) reflection of two superlattices.

**Figure 3.** **Microstructure of Co-doped BaFe$_2$As$_2$ superlattices thin films investigated by TEM. a-b**, HAADF image of the <100> projection of (STO$_{1.2\ nm}$/Co-doped Ba-122$_{13\ nm}$)×24. **c**, Cross-sectional TEM image of the <100> projection of (O-Ba-122$_{3\ nm}$/Co-doped Ba-122$_{13\ nm}$)×24. Arrows indicate nano particle arrays in the O-Ba-122 layer along $ab$-axis. **d**, Planar view TEM image of (O-Ba-122$_{3\ nm}$/Co-doped Ba-122$_{13\ nm}$)×24 showing vertical defects along $c$-axis. Inset shows the high resolution image of vertical defects.

**Figure 4.** **Resistivity as a function of temperature.** $\rho(T)$ from room temperature to below $T_c$. Inset figure is superconducting transition of all films.

**Figure 5.** **$J_c$ as a function of magnetic field. a**, Magnetization $J_c$ as a function of magnetic field at 4.2K with the field applied perpendicular to the plane of all three films. **b**, Transport $J_c$ as a function of magnetic field at 16K with the field applied perpendicular and parallel to the plane of (O-Ba-122$_{3\ nm}$/Co-doped Ba-122$_{13\ nm}$)×24 and Co-doped Ba-122 single layer thin films. **c**, Angular dependence of transport $J_c$ at 4T for all three films at a reduced temperature of $T/T_c$ = ~0.6.

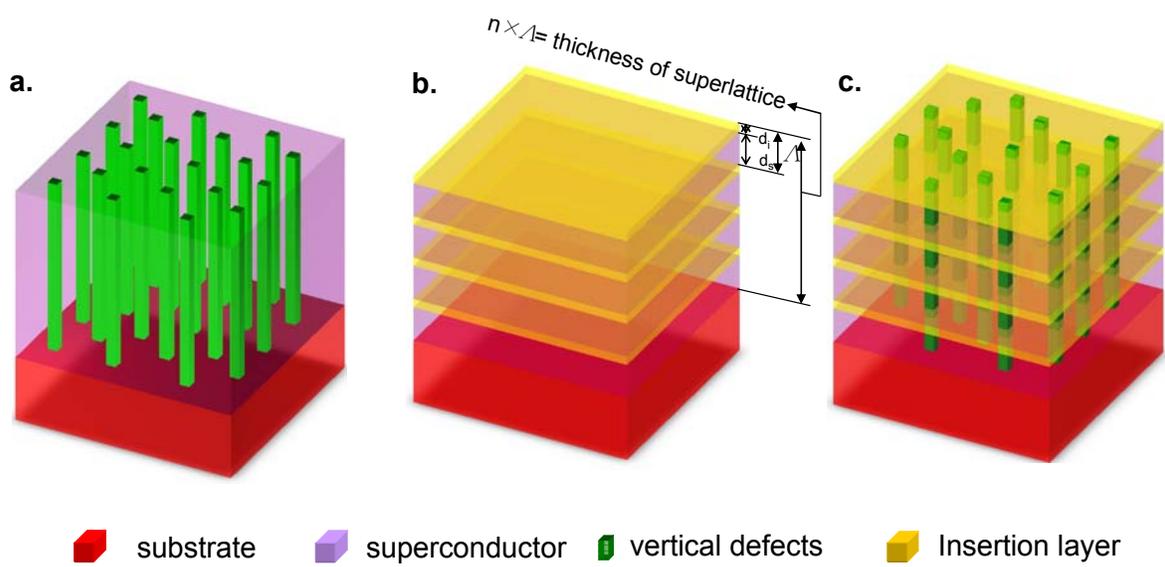

Figure 1

S. Lee *et al*.

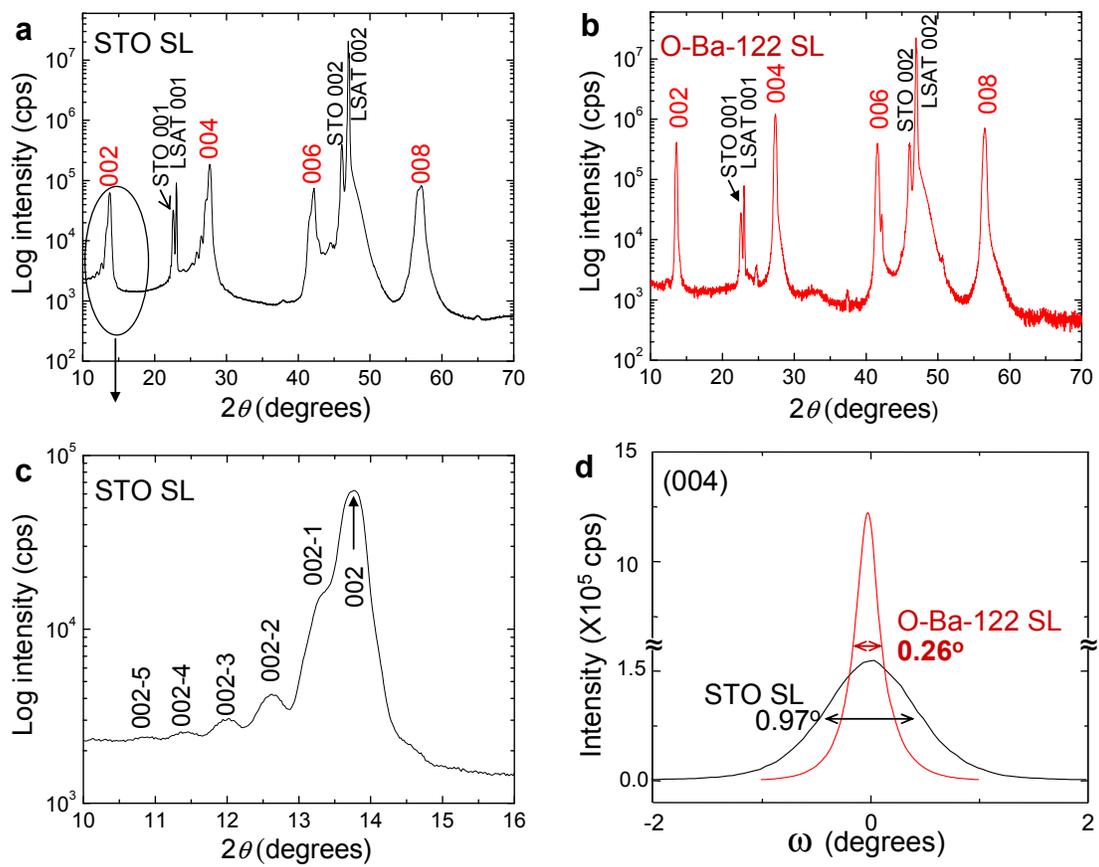

Figure 2

S. Lee *et al*.

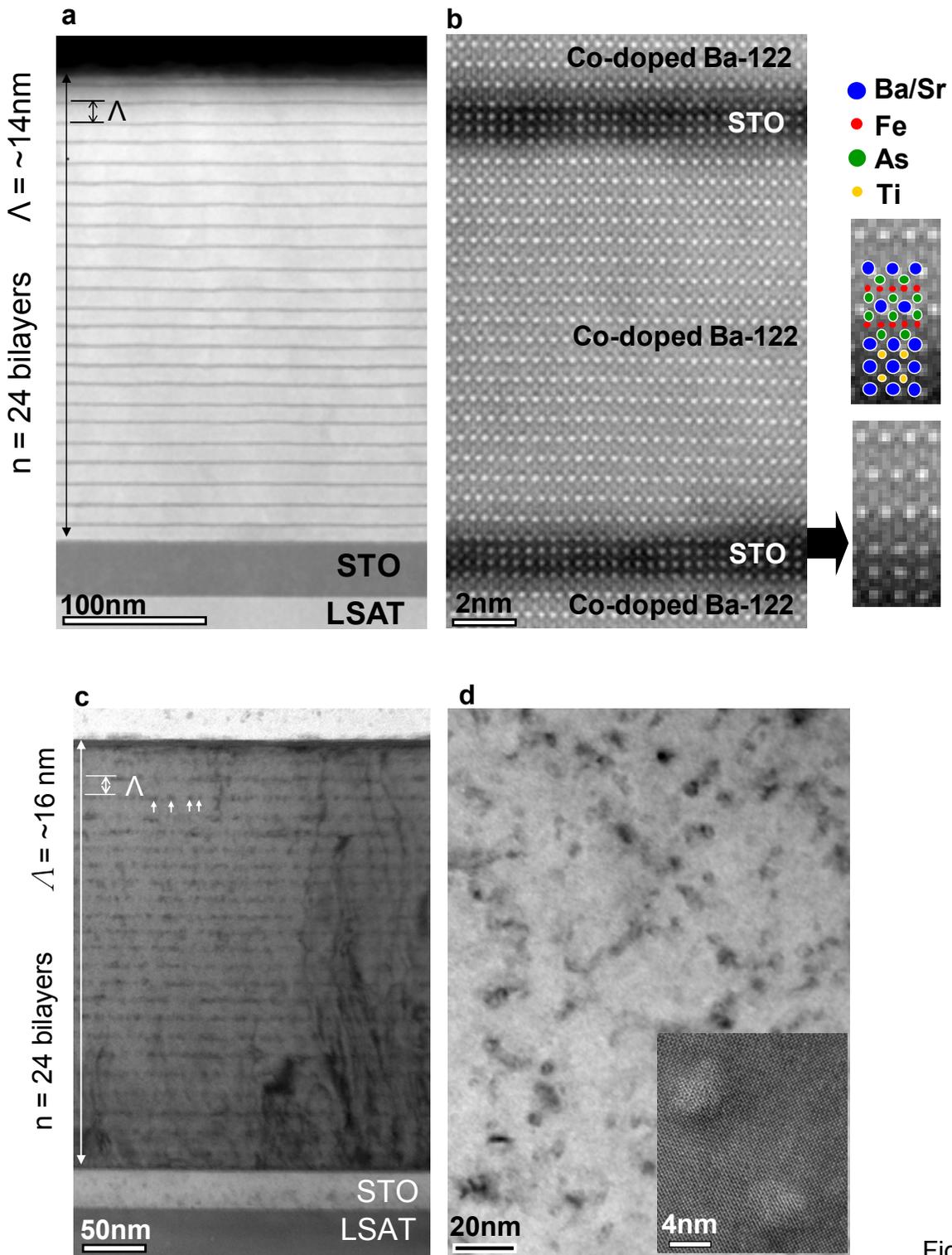

Figure 3

S. Lee *et al*

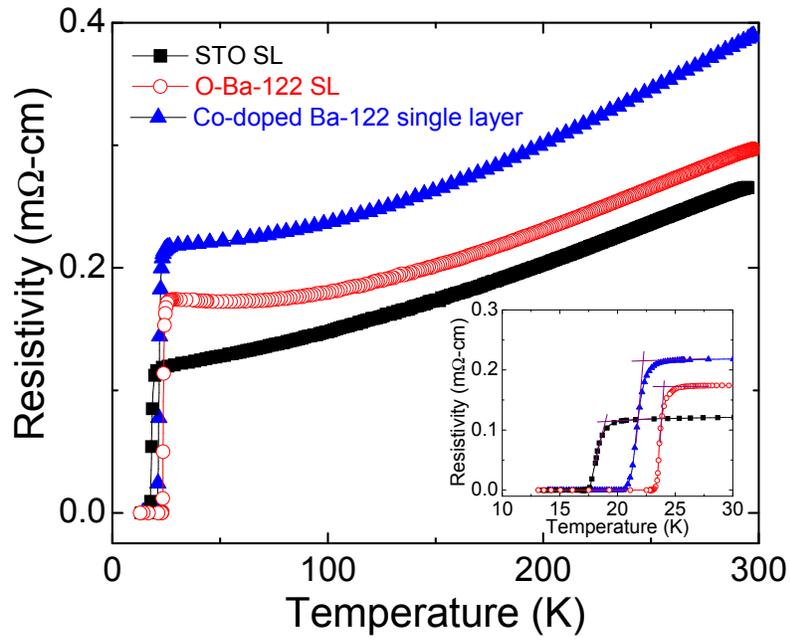

Figure 4

S. Lee *et al*.

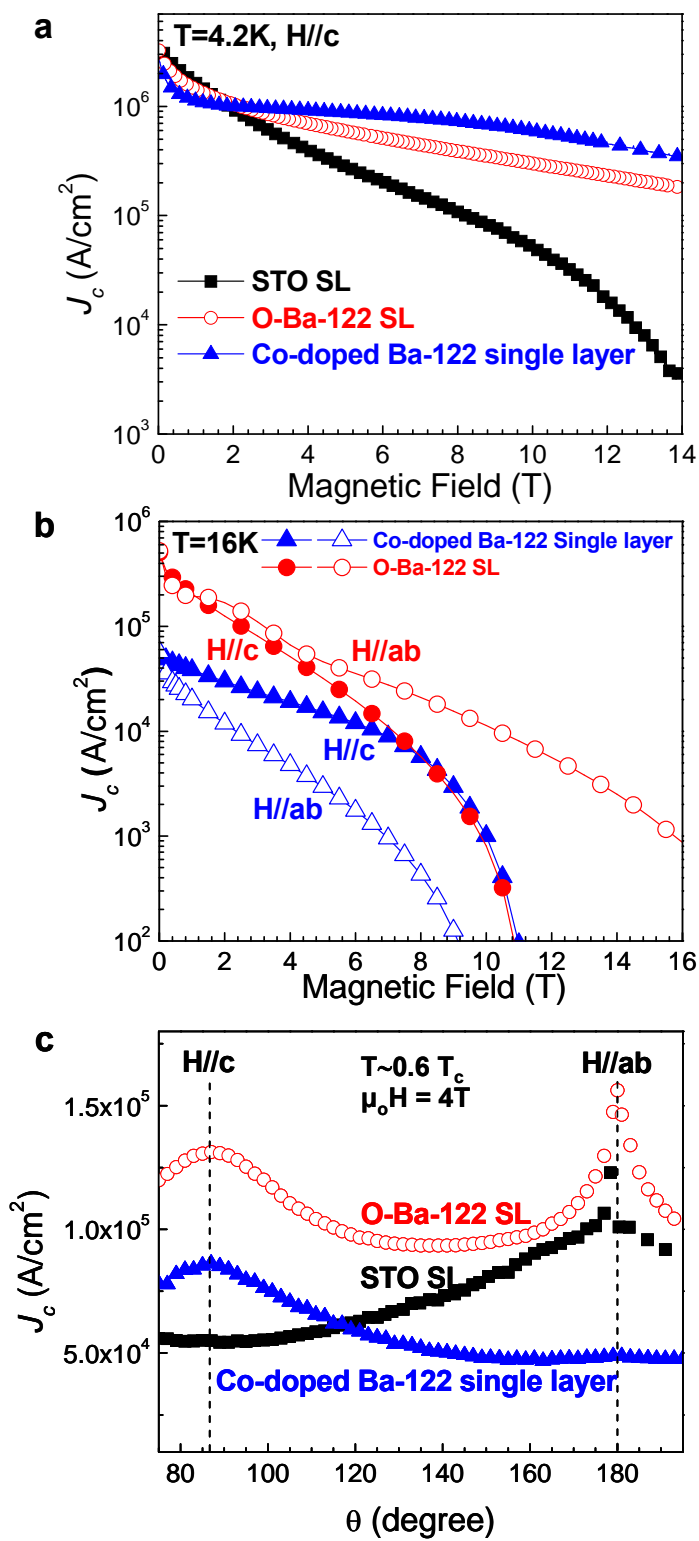

Figure 5

S. Lee et al.

# Supplementary information

## Artificially engineered superlattices of pnictide superconductor

S. Lee[1], C. Tarantini[2], P. Gao[3], J. Jiang[2], J. D. Weiss[2], F. Kametani[2], C. M. Folkman[1], Y. Zhang[3], X. Q. Pan[3], E. E. Hellstrom[2], D. C. Larbalestier[2], and C. B. Eom[1*]

[1]Department of Materials Science and Engineering, University of Wisconsin-Madison, Madison, WI 53706, USA

[2]Applied Superconductivity Center, National High Magnetic Field Laboratory, Florida State University, 2031 East Paul Dirac Drive, Tallahassee, FL 32310, USA

[3]Department of Materials Science and Engineering, The University of Michigan, Ann Arbor, Michigan 48109, USA

Table SI. Crystalline quality and superconducting properties of various ($STO_{x\ nm}$ / Co-doped Ba-122$_{y\ nm}$ and O-Ba-122$_{x\ nm}$ / Co-doped Ba-122$_{y\ nm}$) × n superlattices. As a comparison, the properties of single layer Co-doped Ba-122 thin film are also shown. FWHM of rocking curves of Co-doped Ba-122 (004) reflections are shown as measures of the crystalline quality. $T_{c,\ \rho=0}$, $\Delta T_c$, and magnetization $J_c$ characterize the superconducting properties.

| Co-doped Ba-122 thin films on | FWHM $\Delta\omega$ (°) | $T_{c,\rho=0}$ (K) | $\Delta T_c$ (K) | $J_c$ (4.2K, SF) (MA/cm$^2$) | $J_c$ (4.2K, 10T) (MA/cm$^2$) |
|---|---|---|---|---|---|
| (STO$_{1.2\ nm}$/ Co-doped Ba-122$_{13\ nm}$)X24 | 0.97 | 17.0 | 2.0 | 3.39 | 0.051 |
| (O-Ba-122$_{3\ nm}$/ Co-doped Ba-122$_{13\ nm}$)X16 | 0.27 | 22.3 | 0.9 | 3.73 | 0.296 |
| (O-Ba-122$_{3\ nm}$/ Co-doped Ba-122$_{13\ nm}$)X24 | 0.26 | 22.9 | 1.0 | 3.20 | 0.300 |
| (O-Ba-122$_{3\ nm}$/ Co-doped Ba-122$_{20\ nm}$)X16 | 0.35 | 22.4 | 1.0 | 3.35 | 0.233 |
| (O-Ba-122$_{3\ nm}$/ Co-doped Ba-122$_{20\ nm}$)X24 | 0.31 | 22.5 | 1.3 | 3.44 | 0.263 |
| Single layer Co-doped Ba-122[ref1,2] | 0.55 | 20.5 | 1.9 | 2.90 | 0.596 |

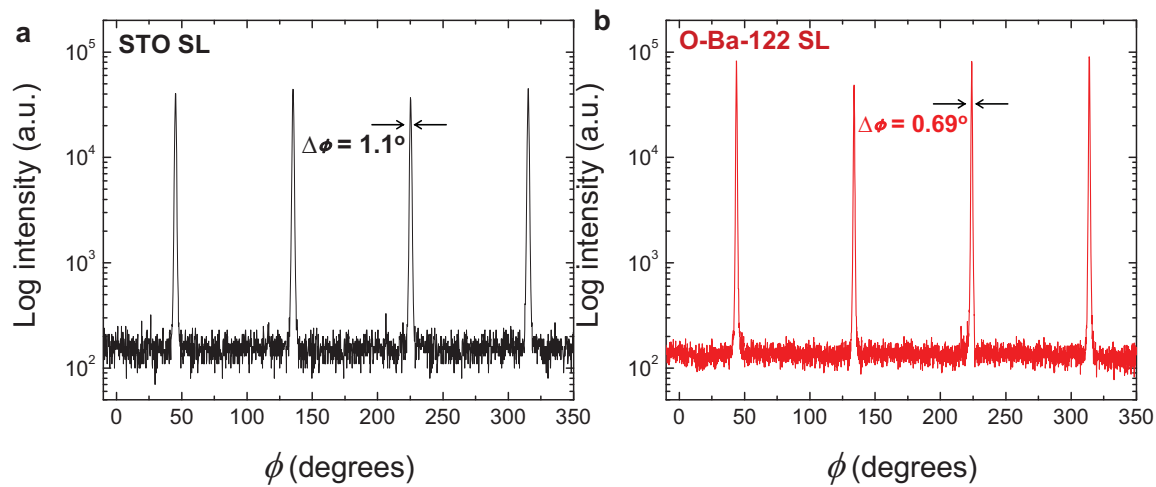

Figure S1. Azimuthal $\phi$ scan and $\Delta\phi$ of the off-axis 112 reflection of **a**, (STO$_{1.2\ nm}$/Co-doped Ba-122$_{13\ nm}$)×24. **b**, (O-Ba-122$_{3\ nm}$/Co-doped Ba-122$_{13\ nm}$)×24 superlattices

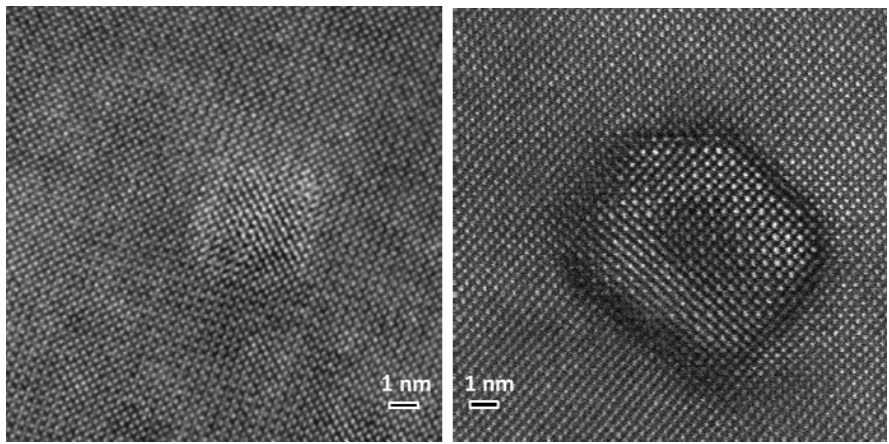

Figure S2. The comparison of (a) the HRTEM and (b) high angle annular dark field (HAADF) images of self-assembled c-axis columnar defect in the same Co-doped Ba122 single layer thin film grown on a SrTiO$_3$/LSAT substrate. It is clear that the HAADF image shows a much better contrast between the 2nd phase and Ba-122 superconducting matrix and the detailed structure can be clearly seen.